\documentclass[runningheads]{llncs}

\usepackage[T1]{fontenc}
\def\doi#1{\href{https://doi.org/\detokenize{#1}}{\url{https://doi.org/\detokenize{#1}}}}
\usepackage{graphicx}
\usepackage{pgfplots, pgfplotstable}

\pgfplotstableread{
X	Y
44.03266144	294
49.69924164	341
49.69924164	341
56.36580658	352
75.13218689	471
75.13218689	471
214.491333	860
239.4852295	902
239.4852295	902
268.4781494	950
269.4779053	952
269.4779053	952
273.4769287	958
308.8016357	1018
308.8016357	1018
394.6806641	1278
411.0100098	1354
411.0100098	1354
44.09932709	334
67.09897614	372
76.43390656	388
100.772438	428
108.4407425	441
164.7861023	570
295.2355957	1063
342.557373	1236
403.8757324	1389
445.8654785	1463
557.8381348	1706
565.8361816	1771
44.13265991	334
68.7995224	375
158.4853516	596
176.1557922	625
210.1557922	722
220.1533508	738
272.8071594	826
285.1374817	896
432.2015686	1345
433.5345764	1347
527.8449097	1469
530.5109253	1473
43.5660019	333
50.8992233	345
55.5658188	353
73.2337112	383
271.240967	891
297.567871	969
413.206299	1361
430.5354	1428
527.511719	1707
530.510986	1712
531.510742	1714
534.176758	1719
556.837891	1804
559.170654	1808
37.3994293	283
138.08432	574
257.940765	924
314.593597	1055
436.897095	1316
440.562866	1323
444.56189	1330
479.886597	1493
482.885864	1498
506.546753	1540
}\datatable
\pgfplotsset{width=10cm,compat=1.9}

\pgfplotstableread{
X	Y
44.03266144	424
49.69924164	433
49.69924164	433
56.36580658	444
75.13218689	457
75.13218689	457
214.491333	659
239.4852295	770
239.4852295	770
268.4781494	818
269.4779053	820
269.4779053	820
273.4769287	826
308.8016357	886
308.8016357	886
394.6806641	1270
411.0100098	1298
411.0100098	1298
44.09932709	464
67.09897614	574
76.43390656	628
100.772438	753
108.4407425	766
164.7861023	898
295.2355957	905
342.557373	1146
403.8757324	1226
445.8654785	1423
557.8381348	1510
565.8361816	1524
44.1326599	464
68.7995224	586
158.485352	626
176.155792	762
210.155792	927
220.153351	868
272.807159	956
285.137482	1012
432.201569	1171
433.534576	1173
527.84491	1264
530.510925	1268
43.5660019	463
50.8992233	509
55.5658188	517
73.2337112	593
271.240967	990
297.567871	1067
413.206299	1316
430.5354	1347
527.511719	1518
530.510986	1523
531.510742	1525
534.176758	1530
556.837891	1650
559.170654	1654
37.3994293	413
138.08432	697
257.940765	943
314.593597	1074
436.897095	1335
440.562866	1342
444.56189	1349
479.886597	1460
482.885864	1465
506.546753	1507
}\datatableTwo
\pgfplotsset{width=10cm,compat=1.9}

\pgfplotstableread{
X	Y
42.99934387	292
70.99891663	339
75.99884033	347
101.3374329	427
156.3491821	519
165.3511047	534
192.023468	618
241.0172882	773
244.0165558	778
252.3478546	828
352.6566772	1157
352.9899292	1158
359.9882202	1170
382.982605	1338
409.6427612	1481
423.9725952	1506
424.3058472	1507
432.9703979	1556
444.3009644	1576
456.9645386	1599
521.2822266	1712
43.69933319	333
100.0359116	642
106.2038956	686
136.7104187	737
264.712738	951
350.6917419	1299
444.6687927	1502
477.3275146	1610
480.9932861	1617
485.9920654	1626
491.6573486	1636
497.6558838	1646
582.3018799	1841
599.6309814	1906
35.03279877	279
97.37561035	653
100.3762512	658
138.7177734	796
262.0397949	1303
293.3654785	1394
295.3649902	1437
296.697998	1439
300.0305176	1445
311.6943359	1499
316.6931152	1543
336.6882324	1578
444.3286438	1891
452.9931946	1956
509.6460266	2036
517.644043	2051
552.96875	2193
555.6347656	2198
565.6323242	2260
44.36598969	334
133.3768005	639
144.3791504	657
144.7125549	658
264.3790588	879
282.0414124	909
407.3441467	1214
412.676178	1223
549.2095337	1667
565.5388794	1832
568.204895	1837
582.2014771	1897
596.1980591	1921
43.56600189	293
137.9115295	710
139.5785522	713
174.9194336	807
187.9222107	829
196.2573242	883
269.910614	1075
301.5695496	1201
304.2355652	1244
314.5663757	1299
315.2328796	1300
322.8976746	1314
324.8971863	1317
353.5568542	1368
487.190918	1712
498.8547363	1732
504.5200195	1742
504.8532715	1743
593.6315918	1976
}\datatableThree
\pgfplotsset{width=10cm,compat=1.9}

\pgfplotstableread{
X	Y
42.99934387	422
70.99891663	469
75.99884033	477
101.3374329	519
156.3491821	501
165.3511047	516
192.023468	536
241.0172882	618
244.0165558	623
252.3478546	637
352.6566772	699
352.9899292	700
359.9882202	712
382.982605	688
409.6427612	735
423.9725952	760
424.3058472	761
432.9703979	666
444.3009644	686
456.9645386	709
521.2822266	712
43.69933319	463
100.0359116	597
106.2038956	607
136.7104187	658
264.712738	652
350.6917419	800
444.6687927	901
477.3275146	958
480.9932861	965
485.9920654	974
491.6573486	984
497.6558838	994
582.3018799	1079
599.6309814	1109
35.03279877	409
97.37561035	553
100.3762512	558
138.7177734	622
262.0397949	867
293.3654785	919
295.3649902	923
296.697998	925
300.0305176	931
311.6943359	951
316.6931152	960
336.6882324	995
444.3286438	1230
452.9931946	1245
509.6460266	1345
517.644043	1360
552.96875	1467
555.6347656	1472
565.6323242	1534
44.36598969	464
133.3768005	613
144.3791504	631
144.7125549	632
264.3790588	879
282.0414124	909
407.3441467	1128
412.676178	1137
549.2095337	1273
565.5388794	1347
568.204895	1352
582.2014771	1377
596.1980591	1456
43.56600189	423
137.9115295	580
139.5785522	583
174.9194336	642
187.9222107	664
196.2573242	718
269.910614	730
301.5695496	783
304.2355652	788
314.5663757	806
315.2328796	807
322.8976746	821
324.8971863	824
353.5568542	875
487.190918	1046
498.8547363	1066
504.5200195	1076
504.8532715	1077
593.6315918	1199
}\datatableFour
\pgfplotsset{width=10cm,compat=1.9}

\usetikzlibrary{pgfplots.groupplots}
\usepackage{listings}
 \usepackage{xcolor}

\colorlet{punct}{red!60!black}
\definecolor{background}{HTML}{EEEEEE}
\definecolor{delim}{RGB}{20,105,176}
\colorlet{numb}{magenta!60!black}

\lstdefinelanguage{json}{
    basicstyle=\normalfont\ttfamily,
    numbers=left,
    numberstyle=\scriptsize,
    stepnumber=1,
    numbersep=8pt,
    showstringspaces=false,
    breaklines=true,
    frame=lines,
    backgroundcolor=\color{background},
    literate=
     *{0}{{{\color{numb}0}}}{1}
      {1}{{{\color{numb}1}}}{1}
      {2}{{{\color{numb}2}}}{1}
      {3}{{{\color{numb}3}}}{1}
      {4}{{{\color{numb}4}}}{1}
      {5}{{{\color{numb}5}}}{1}
      {6}{{{\color{numb}6}}}{1}
      {7}{{{\color{numb}7}}}{1}
      {8}{{{\color{numb}8}}}{1}
      {9}{{{\color{numb}9}}}{1}
      {:}{{{\color{punct}{:}}}}{1}
      {,}{{{\color{punct}{,}}}}{1}
      {\{}{{{\color{delim}{\{}}}}{1}
      {\}}{{{\color{delim}{\}}}}}{1}
      {[}{{{\color{delim}{[}}}}{1}
      {]}{{{\color{delim}{]}}}}{1},
}

\usepackage[english]{babel}

\usepackage{amsmath}
\usepackage[colorlinks=true, allcolors=blue]{hyperref}
\usepackage{color,soul}
\usepackage{booktabs}
\usepackage{verbatim}
\usepgfplotslibrary{external}
\usepackage{caption}
\usepackage{subcaption}

\begin{document}

\title{Embracing AWKWARD!\\ Real-time Adjustment of Reactive Plans Using Social Norms}
\titlerunning{Real-time Adjustment of Reactive Plans Using Social Norms}

\author{Leila Methnani\inst{1}\orcidID{0000-0002-9808-2037} \and
Andreas Antoniades\inst{2}\orcidID{0000-0002-4144-3057} \and
Andreas Theodorou\inst{1}\orcidID{0000-0001-9499-1535}}
\authorrunning{L. Methnani et al.}
\institute{Department of Computing Science, Umeå University, Umeå, Sweden \email{\{leila.methnani,andreas.theodorou\}@cs.umu.se} 
\and Independent scholar \email{andreas.a.antoniades@gmail.com}
}

\maketitle              % typeset the header of the contribution
\begin{abstract}

This paper presents the AWKWARD architecture for the development of hybrid agents in Multi-Agent Systems. AWKWARD agents can have their plans re-configured in real time to align with social role requirements under changing environmental and social circumstances. The proposed hybrid architecture makes use of Behaviour Oriented Design (BOD) to develop agents with reactive planning and of the well-established OperA framework to provide organisational, social, and interaction definitions in order to validate and adjust agents' behaviours. Together, OperA and BOD can achieve real-time adjustment of agent plans for evolving social roles, while providing the additional benefit of transparency into the interactions that drive this behavioural change in individual agents. We present this architecture to motivate the bridging between traditional symbolic- and behaviour-based AI communities, where such combined solutions can help MAS researchers in their pursuit of building stronger, more robust intelligent agent teams. We use DOTA2---a game where success is heavily dependent on social interactions---as a medium to demonstrate a sample implementation of our proposed hybrid architecture.

\end{abstract}

\section{Introduction}

In a Multi-Agent System (MAS) the ability for individual agents to adjust their behaviour when interacting with each other and their environment is critical to the system’s success \cite{dignum2011agents}.
Yet, agents in MAS need to dynamically re-orient their priorities away from their individual---often selfish---goals and towards the system’s collective goals and vice versa as their environment changes. 

One technique used to develop agents in highly dynamic environments is Behaviour-Based Artificial Intelligence (BBAI)\cite{Brooks1991a}. Instead of trying to model the environment, BBAI strictly focuses on the actions that an agent can take and limiting search within a predefined plan for responsive and robust goal-oriented behaviour \cite{Guzel2012,Bryson2001}. 
While this approach does indeed increase the search speed, it reduces the flexibility of the system as it is able to react only to \textit{what its developers have specified}. Moreover, BBAI on its own is insufficient when applied to MAS. It does not account for social interactions between agents or any team work explicitly, and thus fails any consideration of real-world challenges where accounting for social behaviours is required. 

In this paper, we combine BBAI with formal approaches to get the `best of both worlds' by developing a hybrid architecture: {A}gents {W}ith {K}no{W}ledge {A}bout {R}eal-time {D}uties (AWKWARD). We integrated the OperA framework \cite{dignum2004model} with Behaviour-Oriented Design (BOD) \cite{Bryson2001} for their individual and combined strengths in order to produce socially-aware BBAI agents. With OperA, we model the interactions between agents, which contributes towards both governing social behaviour as well as increasing transparency of emerging system behaviour. With BOD, we can build reactive planning agents that are suited to interact within uncertain and dynamic environments. 
We have implemented a `toy example', presented in this paper, for the popular video game DOTA2. Video games have traditionally been used to test AI solutions due to their highly dynamic virtual worlds, which DOTA2 offers as a test bed. Note, we do not consider the specifics of the DOTA2 implementation as our contribution; our focus and contribution is the AWKWARD architecture.
 
The paper is structured as follows: first, we discuss Behaviour Oriented Design and OperA as the backbone of our architecture, outlining the relevant characteristics of each. Next, we introduce the AWKWARD architecture and then a sample implementation of the architecture. In the penultimate section, we look at related work done in normative agents, comparing and contrasting those architectures with our own. Finally, we summarise our contributions and identify future work.

\section{Background}
\subsection{Behaviour Oriented Design}
BOD is a BBAI approach that uses hierarchical representations of an agent's priorities \cite{Bryson2001}. These representations express both the priority of the agent in terms of the goals it needs to achieve, and the contexts in which sets of actions may be applicable \cite{BrysonAgent03}. Another important feature is the usage of the parallel-rooted hierarchy, which allows for the quasi-parallel pursuit of behaviours and a hierarchical structure to aid the design of the agent's behaviour. On each plan cycle, the planner alternates between checking for what is currently the highest-level priority that should be active and then progressing work on that priority. %A plan in BOD consists of the following elements:
Wortham et al. \cite{Wortham2016Instinct} detail the building blocks of a reactive plan in BOD, which are summarised as follows:
\begin{enumerate} 
	\item \textbf{Drive Collection (DC):} The root node of the plan's hierarchy: contains a set of Drives. The DC is responsible for giving attention to the highest priority Drive as at any given cycle only Drive can be active.
	\item \textbf{Drive:} Allows for the design and pursuit of a specific behaviour. Each Drive has its own release condition of one or more Senses. Even when it is not the focus of the planner attention, each Drive maintains its execution state allowing the quasi-parallel execution of multiple drives.
    \item \textbf{Competence:} A self-contained basic reactive plan representing the priorities within the particular plan. Each Competence contains at least one non-concurrent Competence Element (CE). Each of these elements is associated with both a priority relative to the other elements and a context which can perceive and report when that element can execute. The highest-priority action that can be executed will do so when the Competence receives attention.
	\item \textbf{Action Pattern:} Fixed sequences of actions and perceptions used to reduce the design complexity, by determining the execution order in advance.
	\item \textbf{Action:} A possible `doing' of the agent, such the use of an actuator to interact with the environment; i.e. the means of altering the world and self.
	\item \textbf{Sense:} A reading of the world or internal status from a sensor of the agent, such as measuring distance between specified units in the world; i.e. the means of reporting environmental and agent status.
\end{enumerate}
BOD aims to enforce the good-coding practice `Don't Repeat Yourself' by splitting the behaviour into two core modules: the \textit{planner} and the \textit{behaviour library} \cite{BrysonTheodorou2019}. The former reads and `runs' the plan at set intervals. The latter contains the blocks of code used by the two primitive plan elements, Actions and Senses. The rest of the plan elements are textually listed in dedicated files, written in Lisp-like format \cite{Bryson2001}, read by the planner. A plan file contains descriptions of both the plan elements and of the connections between the elements. 

\subsection{OperA}
OperA is an agent organisation framework for the design and development of MAS consisting of three intermingling models \cite{dignum2004model}:
\begin{enumerate} 
\item \textbf{Organisational Model (OM):} Describes objectives and the concerns of the organisation from a social perspective. The development of an OM is approached from a top down perspective, that is, with overarching goals and a means to reach them.

\item \textbf{Social Model:} Outlines the agent's role enactment in the form of social contracts. These social contracts describe what capabilities and responsibilities each role demands. 

\item \textbf{Interaction Model:} Defines interaction agreements between role-enacting agents in the form of interaction contracts. These contracts serve as verification for the fulfilment of interaction agreements between relevant actors specified by organisational objectives as defined in the OM.
\end{enumerate} 

OperA requires that all interactions are expressed as \textit{scene} scripts. A scene is a formal specification defining which \textit{roles} within the organisation partake in the interaction, what \textit{landmarks} from the environment indicate the scene's start, and what resulting signals describe its termination. More importantly, the specification contains a set of \textit{rules} describing the social norms that the participating agents are expected to follow for the scene's full duration. OperA norms use the deontic expressions of obligation, prohibition and permission as a means of describing an agent's social behaviour and further validating whether it satisfies or violates organisational expectations. 

The OperA framework provides a formal specification that depends on organisational structures and global objectives of the organisation as a whole \cite{dignum2004model}. Moreover, OperA offers an interaction model between agents without requiring knowledge of the internal architecture of the individual agent itself; this quality in particular is our primary motivation for selecting OperA as a normative MAS framework. Further motivation is offered in Section 6 where we compare our selection with other related methodologies. 

\section{The AWKWARD Architecture}
\begin{figure}[t]
\centering
 \includegraphics[scale=.45]{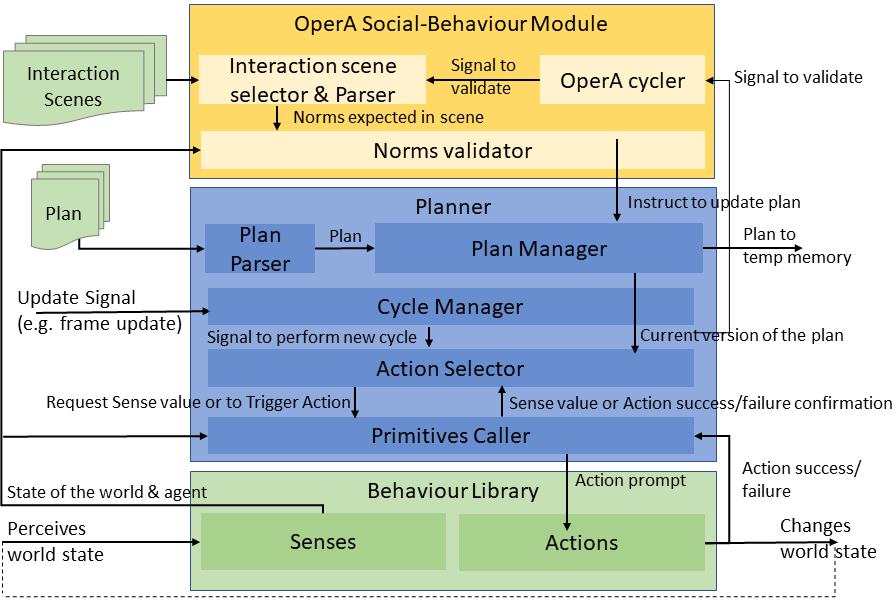}
 \caption{Conceptual diagram of the AWKWARD architecture. The diagram is colour coded; in yellow, representing our System 1, are the parts of the OperA module, in blue are the components of system 2, i.e. the reactive planner; and in green the code components and files shared during execution by multiple agents.}
 \label{fig:arch}
 \centering
\end{figure}

The AWKWARD architecture, depicted in Figure~\ref{fig:arch}, is a hybrid-systems architecture designed for agents operating in multi-agent systems. It consists of three modules, each with a distinct purpose: 1) the reactive planner; 2) the OperA module; and 3) the behaviour library. Our solution, inspired by the dual-process theory presented by \cite{Kahneman2005}, employs a `fast' system 1 and a `slow' system 2 working in tandem for efficient decision making while taking into consideration its wider environmental and social context.

\subsection{The AWKWARD Planner} 
The `fast system' consists of the reactive planner. The planner allows the agent to act upon its intuitions: plans with multiple drives are triggered based on its environmental and internal changes. Each change may enable short-term or long-term goals for the agent to achieve. Reactive planning has the advantage of faster action-selection and the ability to manage dynamic and unpredictable environments \cite{Bryson2001,Brooks1991a}. Most specifically, we use the BOD paradigm due to its proven use in virtual environments, e.g. games \cite{Theodorou2019COG,GaudlFDG13,Brom2006} and simulations \cite{Bryson2007ABM}. BOD, unlike other BBAI approaches, allows the execution of multiple behaviours in pseudo-parallel and has a strong emphasis on modularity and reusability.

BOD plans form a hierarchical tree structure that is traversed from the root to the leaves in order of priority. This order determines the agent’s behaviour given the world circumstance it finds itself in. The hierarchy is predetermined by the plan developer and indicated in the plan. In AWKWARD, plans are written as JSON files. 
At its initiation, the \textit{Plan Parser} component parses the plan to memory, accessed through the \textit{Plan Manager} component, storing the relationships between the plan elements and the hierarchical order of those plan elements. Each agent's DCs can be constructed from the same drive elements, but will differ only in the order of execution, resulting in different expressed behaviours per agent. By initiating all roles with the same plan (i.e. same drive collection hierarchy) and enforcing social norms on agents who violate interaction agreements by explicitly re-prioritising drives, we can shape role- and interaction-dependent plans as needed by the current environment state. 

At set intervals, referred to as \textit{ticks}, the planner's \textit{Cycle Manager} prompts the \textit{Action Selector} component to re-evaluate the agents' perceived conditions to check if a new plan element needs to be executed or the currently running one should continue doing so. This continuous re-evaluation of the current plan elements, called the \textit{plan cycle}, requires access from the planner to the behaviour library. That is, during each cycle, the planner retrieves the sensory inputs in the form of Sense plan elements, and may trigger actuators in the form of Action plan elements. The plan cycle is set on a fixed frequency based on an external update signal; for example, in our toy implementation, the plan cycle is set on every frame update inline with previous implementations of BOD in games \cite{GaudlFDG13}. 

On every tick, the {Action Selector} component retrieves the plan from memory and checks the Drive Collections (DCs) in a hierarchical order. If the conditions of a DC are satisfied, as determined by its corresponding Sense elements, it is executed. The planner then traverses through the drive elements of the DC, checking if they are eligible to be executed or not. These comparisons are done by checking by comparing current sensory reading against a set of preconditions, expressed as sense elements, to determine if the behaviour should be pursued or not by using simple boolean logic. If a drive fires, the planner stops its search at the current tick. This approach of local search enables agents to produce complex behaviours with minimal computational resources as there is no need to explore every possible behaviour at each time \cite{Guzel2012}. If the drive fired is a different one from the previous cycle, then the existing one `pauses.' In other words, at the DC level, different drives can be in different states of execution enabling a quasi-parallel pursuit of multiple behaviours. Instead, the agent focuses---like our system~1 does---on whatever the highest priority behaviour that should be triggered is, e.g. staying alive, instead of unnecessarily checking if lower priority behaviours could also be triggered \cite{Wortham2016Instinct,Bryson2001}. 

\subsection{The OperA Module} 
In AWKWARD, the `slow' system is the OperA module. It validates the social behaviour of the agent and provides direction to the reactive planner upon the completion---either with a success or failure---of a drive's execution. As discussed in section~2, OperA is instantiated with a collection of Interaction Scenes, i.e. formal specification defining which roles within the organisation partake in the defined interaction. Using the senses found in the behaviour library module, the Cycle Manager component prompts the OperA module to check whether a scene has been initiated or terminated. 

While a scene is running, the OperA module verifies that the agent's behaviour fulfils all social obligations that the agent has towards the other agents participating in the same scene. If the agent does not fulfil its obligations, then the OperA module instructs the planner to rearrange the priority of the drives in the currently running Drive Collection. This rearrangement is done by pushing upwards in the hierarchy any drive that corresponds to the desirable social behaviour. The OperA module is informed through formal specifications about which drive should correspond to which social behaviour. It identifies them within the current in-memory version of the plan by using a string equality operation. When the adjustment of drives is done, the new plan is stored in the system's memory, overwriting the old version. The OperA module checks if the re-prioritisation has produced behaviour that falls within the social and organisational norms the agent needs to comply with. If not, the OperA module continues to adjust the plan's drives further until the expected social behaviour is achieved. For this social norm validation, OperA uses the same senses as the planner component does to check which plan elements should be executed. 

In its past implementations, a single instance of OperA operated at a global level, i.e. it was responsible for agents and interactions in the model. In AWKWARD, each agent contains its own local instance of OperA; i.e. each AWKWARD agent only checks its own behaviour in its current interaction scene. An advantage of this approach is that it allows us, at least in larger environments, to keep track of the multiple localised interaction scenes that the agent can be in simultaneously.

The use of a localised instance of OperA takes inspiration from Dennett’s description of social constructs, where local efforts are made by agents to ``steer their part of the whole contraption while [remaining] blissfully ignorant of the complexities on which the whole system depends'' \cite{dennett2020age}. That is to say, each agent does not need to have any conceivable notion of the global system’s intricacies, but rather a partial comprehension which may suffice for competence. In practical terms, as demonstrated in our toy example that is further discussed in section~4, by having local copies of OperA carried with each agent also means that an AWKWARD agent is not constrained to interacting with other AWKWARD agents only. Rather, the agent can consider and interact with any agent architecture, as well as humans. Moreover, an AWKWARD agent does not require perfect knowledge of all agents in the same environment---which is impossible to maintain in certain scenarios, e.g. competitive games---to continuously adjust and moderate its behaviour within a defined socio-organisational role.

\subsection{Behaviour Library} 

While each agent has its own individual plan structure, a single instance of the behaviour library can be shared across all agents. It is the collection of all possible primitives (i.e. Senses and Actions) as discussed in section~2.1. It is accessed through global function calls in the primitives' tick functions. The behaviour library must therefore maintain a direct pointer to the agent that makes any particular function call in order to return values appropriate to the agent that requested them. By having a single behaviour library for all agents, we not only enable code reusability, but also reduce memory footprint of the agents and support the hosting of a remote behaviour library that runs complex actions and senses \cite{BrysonTheodorou2019}. 

As discussed above, the behaviour library is accessible by both the planner and the OperA modules as they each provide control over the agent's means of environmental perception, internal status, and any actuators available. This decision to reuse the same Senses enables code reusability but also constrains OperA's knowledge to that of the agent, which it uses for the action-selection process. Arguably, this is also a more realistic implementation of Kahneman's dual-system theory \cite{Kahneman2005}: our system 1 acts reactively and system 2 acts deliberately, but the embodiment with its sensors and actuators, i.e. behaviour library, is shared by both systems.

\section{Implementation in DOTA2}

We developed a `toy example' in DOTA2, a popular game characterised by its extremely steep learning curve and complex emerging behaviour through the interactions of the different actors between themselves and their environment. We selected this game due to its complexity, inherent need for a MAS approach, recognition in the AI research community, and access to a public API for AI researchers\footnote{Available at: \url{https://developer.valvesoftware.com/wiki/Dota\_Bot\_Scripting}}. Moreover, DOTA2 has been used in the past by the AI community; OpenAI \cite{berner2019dota} developed agents that were able to outperform 99\% of the DOTA2 player base. Our interest, unlike OpenAI, extends beyond the scope of the individual agent and creating human-beating agents. Instead, we used this highly dynamic environment to demonstrate how AWKWARD can be implemented and have BBAI agents working together in a team by using OperA to model and validate their social interactions.

\subsection{DOTA2}
In the game, two opposing teams of five players (i.e. agents) navigate through terrain, striving to destroy a structure in their enemy's base known as the Ancient, while also defending their own.  
There are five roles---or positions---to fill per team. Each team agent is a \textit{hero} that is assigned a position complementary to their given skill-set. For instance, a hero with healing capabilities can fulfil the responsibilities of a \textit{Position 5} role, which includes supporting a hero in the \textit{Position 1} role. In the early game, the Position 1 role is one of the weakest members. As the match progresses, Position 1 typically becomes the strongest on the team. However, to reach that state, a social norm exercised in most games involves giving Position 1 priority, at least in certain scenes, to perform the self-advancing activity known as \textit{farming}. However, given specific circumstances, supporting roles may break that norm to farm in favour of advancing themselves. Hence, it is important to alternate strategies between pursuing behaviours for the benefit of others on the team and for your own personal performance. The resulting team of agents demonstrate emergent behaviour in the arena as they interact. This emergent behaviour can be difficult to perfectly model or even understand from the observer’s perspective.

\subsection{The Reactive Planner Module and Behaviour Library}
As we discussed in the previous section, AWKWARD consists of a planner module and a behaviour library. In order for each agent to behave in compliance with its assigned role, its dedicated plan must consist of a distinctly ordered collection of drives (i.e. DCs). In our DOTA2 implementation, the plan is described in a JSON string that is parsed by the planner. We synchronised the planner update frequency with the game's internal execution update; i.e. the planner ticks the DC to check its drives on every frame update. In our toy example, we implemented a sample DC with multiple drives. 

One of our implemented drives is the \texttt{DE-FarmLane}, represented in Formalism~\ref{eq:farm}. The drive prompts the behaviour of seeking out enemy units called \textit{creeps} in the hero's assigned lane in the environment and striking them when they are low on health in order to achieve what is referred to as a \textit{last hit}. Last hits result in a kill and a gold bounty for the hero to collect. Gold is the currency used to purchase items in-game providing heroes with added attributes and abilities in battle. For example, a \textit{healing salve} can be purchased and consumed to aid in health regeneration. 

The option for health regeneration is captured by Formalism~\ref{eq:heal}. This drive is executed in response to the agent's internal state: a measure of low health, as defined by the agent designer. It is important to note that two drives such as the external state of \textit{farm time} in \texttt{DE-FarmLane} and internal state of \textit{low health} in \texttt{DE-Heal}, could be true at the same time. In our current implementation, which behaviour is expressed is determined by plan structure; namely the order of executing drives. For example, if \texttt{DE-Heal} is prioritised over \texttt{DE-FarmLane}, then it will execute for as long as the world satisfies its \textit{low health} condition, or until a drive of higher priority is able to run. For instance, if the agent has low health while simultaneously taking damage the plan designer might prioritise a behaviour that more urgently involves retreating.  

\begin{eqnarray} \label{eq:farm}
\resizebox{0.5\hsize}{!}{$\mbox{farm time} \Rightarrow
\left\uparrow
\left\langle \begin{array}{r@{ \Rightarrow }l}
	\mbox{laning phase ended?} & \mbox{ {\em goal}}\\
	\mbox{creep can be last hit?}  & \mbox{ lastHitCreep} \\
	\mbox{creep wave far?}  & \mbox{ goToCreepWave} \\
	  & \mbox{ goToAssignedLane} 
	\end{array}
	 \right\rangle \right.$}
\end{eqnarray}

\begin{eqnarray} \label{eq:heal}
\resizebox{0.5\hsize}{!}{$\mbox{low health} \Rightarrow
\left\uparrow
\left\langle \begin{array}{r@{ \Rightarrow }l}
	\mbox{full health?} & \mbox{ {\em goal}}\\
	\mbox{healing ability?}  & \mbox{ use healing ability} \\
	\mbox{healing item?}  & \mbox{ use healing item} \\
	\mbox{enough gold?}  & \mbox{ buy healing item} \\
	  & \mbox{ retreat} 
	\end{array}
	 \right\rangle \right.$}
\end{eqnarray}

The drive element \texttt{DE-Retreat}, represented in Formalism~\ref{eq:retreat}, prompts the behaviour of seeking refuge in the occurrence of declining health. This drive, as it is currently implemented, only fires when the hero is below 30\% health and directs the hero towards their home base where their \textit{fountain} resides and provides protection while regenerating health.

\begin{eqnarray} \label{eq:retreat}
\resizebox{0.5\hsize}{!}{$\mbox{under attack} \Rightarrow
\left\uparrow
\left\langle \begin{array}{r@{ \Rightarrow }l}
	\mbox{full health?} & \mbox{ {\em goal}}\\
	\mbox{low health AND taking damage? }  & \mbox{ retreat} \\
	\end{array}
	 \right\rangle \right.$}
\end{eqnarray}

Drives may fire Competences or Action Patterns. The competences consist of an ordered list of competence elements that fire other competences or action patterns. The action patterns consist of a sequence of one or more actions---a primitive element found in the behaviour library along with senses. For instance, Formalism~\ref{eq:farm} depicts \texttt{DE-FarmLane} and lists \texttt{lastHitCreep} as a competence element that fires the action pattern shown in Formalism~\ref{eq:lasthit}. The action patterns consist of \texttt{selectTarget} followed by \texttt{rightClickAttack}, which are defined in the behaviour library.

\begin{eqnarray} \label{eq:lasthit}
\langle selectTarget \rightarrow rightClickAttack \rangle	
\end{eqnarray}

An agent's individual behavioural desire is shaped by how its plan element are arranged and executed. Consider again the drive element \texttt{DE-FarmLane}, which encourages a hero to individually collect as much bounty from last hitting lane creeps as possible. All heroes have this drive in their plans, and while it is a common desire for all members of the organisation to maximise their individual farm, it would not benefit the collective team if all members were to continuously act selfishly. 
For instance, in the early game a hero with the role of Position 1 is often vulnerable and weak with little of the gold needed to mature their abilities and grow their arsenal. It is therefore the duty of the Position 5 role to sacrifice farm for the sake of their allied member who requires it more at this early stage. Social interactions become particularly interesting and important for us to capture here. This is also where the OperA framework shines, as it enables the definition of social role assignments and interaction agreements for the advancement of the whole team unit.

\subsection{The OperA Module Implementation}
We use social interactions to alter the order in which a reactive agent's drives are fired. Accomplishing this real-time adjustment should demonstrate successful expression of social behaviour. That is, altering selfish priorities for the collective good of the organisation the individual hero is a member of. The OperA module requires a record of the relevant members and their associated roles within the society. Currently, each hero's role assignment also determines their lane assignment in the environment, which further characterises their ``right''.

For example, Table~\ref{tab:position1} outlines the role of \textit{Position~1}, indicating that their ultimate objective is to ensure their team's victory, which can only be done by collecting enough items to become powerful. Their sub-objectives are therefore to farm as much as possible, buying items with earned gold from farm. They have the right to do so in their assigned zone: the \textit{safe lane}---named as such due to its proximity to safety zones. An example norm that applies to a Position 1 hero is the obligation to farm enemy creeps when they are close by. In contrast, Table~\ref{tab:position5} shows the role table for Position 5, who is not permitted to farm while Position 1 is nearby. This sacrifice is to ensure Position 1 gains enough gold to quickly advance their role and carry the team in the later phases of the game.

\begin{table}[]
\centering
  \caption{Position 1 role defined using the OperA framework}
  \label{tab:position1}
  \begin{tabular}{ll}\toprule
  \textbf{role id} & Position 1 \\ \midrule
  \textbf{objectives} & carry team to victory\\ 
  \hline
  \textbf{sub-objectives} & farm and buy items\\
  \hline
  \textbf{rights} & high priority in safe lane \\
 \hline
 \textbf{rules} &  \begin{tabular}{@{}l@{}}\texttt{IF} enemy
 creep around \texttt{THEN} \\ \qquad \texttt{OBLIGED} to farm
 \end{tabular} \\\bottomrule
 \end{tabular}
 \end{table}

\begin{table}[]
\centering
  \caption{Position 5 role defined using the OperA framework}
  \label{tab:position5}
  \begin{tabular}{ll}\toprule
  \textbf{role id} & Position 5 \\ \midrule
  \textbf{objectives} & support team to victory\\ 
  \hline
  \textbf{sub-objectives} & heal allies and take fights\\
  \hline
  \textbf{rights} & low priority in safe lane \\
 \hline
 \textbf{rules} &  \begin{tabular}{@{}l@{}}\texttt{IF} Position 1 nearby \texttt{THEN} \\ \qquad \texttt{NOT PERMITTED} to farm
 \end{tabular} \\\bottomrule
 \end{tabular}
 \end{table}

In the OperA module, relevant norms are constructed by parsing JSON strings with their descriptions. Each norm has a name identifier, an associated behaviour, and a deontic operator. The behaviour corresponds to the suitable drive element of the agent's planner. The deontic operators we focus on for this implementation are \texttt{NOT PERMITTED} and \texttt{OBLIGED}. The \texttt{PERMITTED} operator is a softer norm that induces no change to the plan in the current implementation. A norm also has a reference to an assigned agent's plan. It is validated by checking the agent's active drive against the expected (norm) behaviour, and whether it is permissible or required within a given circumstance. If a norm is violated, sanctions should be applied. In our implementation, a norm can alter the plan in response to a violation. Recall that our planner implementation allows for the removal and insertion of drive elements at runtime. 

\vspace{-2pt}
While norms can---and should---be associated with the individual agent, what is most interesting for our purposes is the use of norms to characterise interaction scenes between agents. An OperA interaction scene has a unique name identifier, a list of roles involved in the scene, a list of landmarks that indicate the start of a scene, a list of results that indicate the end of a scene, and the rules that indicate the norms that constrain the agents' behaviours such that they remain within the social expectations of the team. In our implementation of scene objects, the landmarks and results correspond to sense primitives that were re-used from the reactive plan elements, as do the rule conditions that determine the appropriate norm to apply in the scene.

\begin{table}[t]
\centering
  \caption{Interaction Scene for Priority Lane Farming}
  \label{tab:scene}
  \begin{tabular}{ll}\toprule
  \textbf{scene} & priority lane farming \\ \midrule
  \textbf{roles} & carry and support\\ 
  \hline
  \textbf{landmarks} & partner and creeps nearby\\
  \hline
  \textbf{results} & partner not nearby\\
  \hline
  \textbf{rules} & \begin{tabular}{@{}l@{}}\texttt{IF} highest priority around \texttt{THEN} \\ \qquad \texttt{OBLIGED} to farm \\ \texttt{ELSE} \\ \qquad \texttt{NOT PERMITTED} to farm
 \end{tabular} \\\bottomrule
 \end{tabular}
 \end{table}

Consider again the agent's \texttt{DE-FarmLane} (Formalism~\ref{eq:farm}). This drive is fired when the senses \texttt{isFarmingTime} and \texttt{isSafeToFarm} return true. We have defined farming time as the early game (approximately the first 10 minutes) and safety corresponds to the fight activity and whether any enemy heroes are threatening the agent's farm ability. For the purposes of this demonstration, we assume it is always safe to farm and the sense will return true. The competence elements that the farming drive consists of checks conditions in a prioritised order: if the agent is not at the creep wave, they must move there. (Unlike action patterns which always execute in the same order, competences will skip this element if the condition returns false). If the agent is already at the creep wave, then the next condition check is whether any enemy creep around can be last hit. A creep can be last hit when their health is lower than the hero's attack damage and is located within attack range. If this condition is true, the agent will select the appropriate target and attempt to land the last hit. Notice that the drive is entirely self-directed and lacks any social consideration. That is, each agent, while expressing farming behaviour, will pay no mind to whether they have the highest priority around or not. An agent pair farming in the same lane is not optimal behaviour from a social perspective. The agents must abide by social norms for this given circumstance, and OperA can facilitate this interaction using scene scripts. The result is preservation of interaction agreements and altered plans that better suit each agent's role requirements.
  
While this farm priority check is a simple example that can just as easily be incorporated into each agent's plan, we argue that it will become limiting and wasteful over the course of the game. We observed that when social awareness was incorporated into an individual agent's plan, the structure not only became longer to read and more complicated to understand, but also resulted in increased idle time while an agent was attempting to express farming behaviour. When an agent of higher priority is farming, the agent should not continue to attempt to farm, but should instead fire a different drive for productivity. This coordination is best handled by an organisational/social-aware structure, like OperA. We argue that the individual agent itself should not have too many intricate details about interactions, especially considering the long-time horizons of a game like DOTA2 where social behaviour itself is expected to shift along with the various game phases. In fact, OperA scene scripts very nicely accommodate this game attribute. If behaviour can be altered by OperA, then the reactive planners become simpler to construct. The complexity is captured and described by the OperA interaction models.

\section{Results}

In this paper, we are exclusively concerned with developing a system that can alter agent plans in response to social interactions, regardless of whether other entities in the society have a similar architecture or not. Hence, we designed the evaluation of our sample implementation to reflect that.
For the scope of this project, we focus only on the first 10 minutes of the game; within this time in particular, farming priority is important in terms of behaviour adjustment due to expected social norms. 

The interaction under evaluation is described by the Priority Farm scene as defined by an interaction scene object. The particular behaviour that is expected is for the Position 5 agent---the AWKWARD bot---to give up its own farm for the benefit of the ally of higher priority in the same lane. In this case, the relevant ally is the Position 1 agent---the default bot. 

\begin{figure}
    \centering
    \begin{tikzpicture}
\begin{groupplot}[
    group style={
        group name=my plots,
        group size=2 by 1,
        ylabels at=edge left,
        horizontal sep=1.5cm
    },
    footnotesize,
    width=6cm,
    height=6cm,
    tickpos=left,
    xtick align=outside,
    ytick align=outside,
    xmin=0, xmax=610,
    ymin=100, ymax=2500,
    xtick={100,200,300,400,500,600},
    ytick={400,800,1200,1600, 2000, 2400},
    legend pos=north west,
    ymajorgrids=true,
    grid style=dashed,
    % ylabel=$Gold$,
    xlabel=$Time$ $(s)$,
]
\nextgroupplot[title={No OperA Intervention}, ylabel={$Gold$}]
\addplot [thick, blue] table[
    y={create col/linear regression={y=Y}}
] 
{\datatable};
\addplot [thick, orange] table[
    y={create col/linear regression={y=Y}}
] 
{\datatableTwo};
\addplot[
    color=blue,
    mark size=0.75,
    only marks,
    ]
    coordinates {
(44.032661437988, 294)
(49.699241638184, 341)
(49.699241638184, 341)
(56.36580657959, 352)
(75.132186889648, 471)
(75.132186889648, 471)
(214.49133300781, 860)
(239.48522949219, 902)
(239.48522949219, 902)
(268.47814941406, 950)
(269.47790527344, 952)
(269.47790527344, 952)
(273.47692871094, 958)
(308.80163574219, 1018)
(308.80163574219, 1018)
(394.6806640625, 1278)
(411.01000976562, 1354)
(411.01000976562, 1354)
(44.099327087402, 334)
(67.098976135254, 372)
(76.433906555176, 388)
(100.77243804932, 428)
(108.44074249268, 441)
(164.78610229492, 570)
(295.23559570312, 1063)
(342.55737304688, 1236)
(403.87573242188, 1389)
(445.86547851562, 1463)
(557.83813476562, 1706)
(565.83618164062, 1771)
(44.132659912109, 334)
(68.799522399902, 375)
(158.4853515625, 596)
(176.15579223633, 625)
(210.15579223633, 722)
(220.15335083008, 738)
(272.80715942383, 826)
(285.13748168945, 896)
(432.20156860352, 1345)
(433.53457641602, 1347)
(527.84490966797, 1469)
(530.51092529297, 1473)
(43.5660019, 333)
(50.8992233, 345)
(55.5658188, 353)
(73.2337112, 383)
(271.240967, 891)
(297.567871, 969)
(413.206299, 1361)
(430.5354, 1428)
(527.511719, 1707)
(530.510986, 1712)
(531.510742, 1714)
(534.176758, 1719)
(556.837891, 1804)
(559.170654, 1808)
(37.3994293, 283)
(138.08432, 574)
(257.940765, 924)
(314.593597, 1055)
(436.897095, 1316)
(440.562866, 1323)
(444.56189, 1330)
(479.886597, 1493)
(482.885864, 1498)
(506.546753, 1540)
    };
\addplot[color=orange,
    mark size=0.75,
    only marks,]     
    coordinates {
(44.032661437988, 424)
(49.699241638184, 433)
(49.699241638184, 433)
(56.36580657959, 444)
(75.132186889648, 457)
(75.132186889648, 457)
(214.49133300781, 659)
(239.48522949219, 770)
(239.48522949219, 770)
(268.47814941406, 818)
(269.47790527344, 820)
(269.47790527344, 820)
(273.47692871094, 826)
(308.80163574219, 886)
(308.80163574219, 886)
(394.6806640625, 1270)
(411.01000976562, 1298)
(411.01000976562, 1298)
(44.099327087402, 464)
(67.098976135254, 574)
(76.433906555176, 628)
(100.77243804932, 753)
(108.44074249268, 766)
(164.78610229492, 898)
(295.23559570312, 905)
(342.55737304688, 1146)
(403.87573242188, 1226)
(445.86547851562, 1423)
(557.83813476562, 1510)
(565.83618164062, 1524)
(44.1326599, 464)
(68.7995224, 586)
(158.485352, 626)
(176.155792, 762)
(210.155792, 927)
(220.153351, 868)
(272.807159, 956)
(285.137482, 1012)
(432.201569, 1171)
(433.534576, 1173)
(527.84491, 1264)
(530.510925, 1268)
(43.5660019, 463)
(50.8992233, 509)
(55.5658188, 517)
(73.2337112, 593)
(271.240967, 990)
(297.567871, 1067)
(413.206299, 1316)
(430.5354, 1347)
(527.511719, 1518)
(530.510986, 1523)
(531.510742, 1525)
(534.176758, 1530)
(556.837891, 1650)
(559.170654, 1654)
(37.3994293, 413)
(138.08432, 697)
(257.940765, 943)
(314.593597, 1074)
(436.897095, 1335)
(440.562866, 1342)
(444.56189, 1349)
(479.886597, 1460)
(482.885864, 1465)
(506.546753, 1507)
    };
    \nextgroupplot[title={OperA Intervention}]
\addplot [thick, blue] table[
    y={create col/linear regression={y=Y}}
] 
{\datatableThree};
\addplot [thick, orange] table[
    y={create col/linear regression={y=Y}}
] 
{\datatableFour};
\addplot[color=blue,
    mark size=0.75,
    only marks,] coordinates{
(42.99934387207, 292)
(70.998916625977, 339)
(75.998840332031, 347)
(101.33743286133, 427)
(156.34918212891, 519)
(165.35110473633, 534)
(192.02346801758, 618)
(241.01728820801, 773)
(244.01655578613, 778)
(252.34785461426, 828)
(352.65667724609, 1157)
(352.98992919922, 1158)
(359.98822021484, 1170)
(382.98260498047, 1338)
(409.64276123047, 1481)
(423.97259521484, 1506)
(424.30584716797, 1507)
(432.97039794922, 1556)
(444.30096435547, 1576)
(456.96453857422, 1599)
(521.2822265625, 1712)
(43.699333190918, 333)
(100.03591156006, 642)
(106.20389556885, 686)
(136.71041870117, 737)
(264.71273803711, 951)
(350.69174194336, 1299)
(444.66879272461, 1502)
(477.32751464844, 1610)
(480.99328613281, 1617)
(485.99206542969, 1626)
(491.65734863281, 1636)
(497.65588378906, 1646)
(582.30187988281, 1841)
(599.63098144531, 1906)
(35.03279876709, 279)
(97.375610351562, 653)
(100.3762512207, 658)
(138.7177734375, 796)
(262.03979492188, 1303)
(293.36547851562, 1394)
(295.36499023438, 1437)
(296.69799804688, 1439)
(300.03051757812, 1445)
(311.6943359375, 1499)
(316.69311523438, 1543)
(336.68823242188, 1578)
(444.32864379883, 1891)
(452.99319458008, 1956)
(509.64602661133, 2036)
(517.64404296875, 2051)
(552.96875, 2193)
(555.634765625, 2198)
(565.63232421875, 2260)
(44.365989685059, 334)
(133.37680053711, 639)
(144.37915039062, 657)
(144.71255493164, 658)
(264.37905883789, 879)
(282.04141235352, 909)
(407.34414672852, 1214)
(412.67617797852, 1223)
(549.20953369141, 1667)
(565.53887939453, 1832)
(568.20489501953, 1837)
(582.20147705078, 1897)
(596.19805908203, 1921)
(43.56600189209, 293)
(137.91152954102, 710)
(139.57855224609, 713)
(174.91943359375, 807)
(187.92221069336, 829)
(196.25732421875, 883)
(269.91061401367, 1075)
(301.56954956055, 1201)
(304.23556518555, 1244)
(314.56637573242, 1299)
(315.23287963867, 1300)
(322.89767456055, 1314)
(324.8971862793, 1317)
(353.55685424805, 1368)
(487.19091796875, 1712)
(498.85473632812, 1732)
(504.52001953125, 1742)
(504.85327148438, 1743)
(593.63159179688, 1976)
        };
\addplot[color=orange,
    mark size=0.75,
    only marks,] coordinates{
((42.99934387207, 422)
(70.998916625977, 469)
(75.998840332031, 477)
(101.33743286133, 519)
(156.34918212891, 501)
(165.35110473633, 516)
(192.02346801758, 536)
(241.01728820801, 618)
(244.01655578613, 623)
(252.34785461426, 637)
(352.65667724609, 699)
(352.98992919922, 700)
(359.98822021484, 712)
(382.98260498047, 688)
(409.64276123047, 735)
(423.97259521484, 760)
(424.30584716797, 761)
(432.97039794922, 666)
(444.30096435547, 686)
(456.96453857422, 709)
(521.2822265625, 712)
(43.699333190918, 463)
(100.03591156006, 597)
(106.20389556885, 607)
(136.71041870117, 658)
(264.71273803711, 652)
(350.69174194336, 800)
(444.66879272461, 901)
(477.32751464844, 958)
(480.99328613281, 965)
(485.99206542969, 974)
(491.65734863281, 984)
(497.65588378906, 994)
(582.30187988281, 1079)
(599.63098144531, 1109)
(35.03279876709, 409)
(97.375610351562, 553)
(100.3762512207, 558)
(138.7177734375, 622)
(262.03979492188, 867)
(293.36547851562, 919)
(295.36499023438, 923)
(296.69799804688, 925)
(300.03051757812, 931)
(311.6943359375, 951)
(316.69311523438, 960)
(336.68823242188, 995)
(444.32864379883, 1230)
(452.99319458008, 1245)
(509.64602661133, 1345)
(517.64404296875, 1360)
(552.96875, 1467)
(555.634765625, 1472)
(565.63232421875, 1534)
(44.365989685059, 464)
(133.37680053711, 613)
(144.37915039062, 631)
(144.71255493164, 632)
(264.37905883789, 879)
(282.04141235352, 909)
(407.34414672852, 1128)
(412.67617797852, 1137)
(549.20953369141, 1273)
(565.53887939453, 1347)
(568.20489501953, 1352)
(582.20147705078, 1377)
(596.19805908203, 1456)
(43.56600189209, 423)
(137.91152954102, 580)
(139.57855224609, 583)
(174.91943359375, 642)
(187.92221069336, 664)
(196.25732421875, 718)
(269.91061401367, 730)
(301.56954956055, 783)
(304.23556518555, 788)
(314.56637573242, 806)
(315.23287963867, 807)
(322.89767456055, 821)
(324.8971862793, 824)
(353.55685424805, 875)
(487.19091796875, 1046)
(498.85473632812, 1066)
(504.52001953125, 1076)
(504.85327148438, 1077)
(593.63159179688, 1199)
        };
\end{groupplot}
\end{tikzpicture}
    \caption{(Left) Similar rate of gold acquisition between Position 5 (orange) and Position 1 (blue) DOTA2 agents when OperA makes no alterations to Position 5's plan during the Priority Farm interaction scenes. (Right) Diverging rate of gold acquisition between Position 5 and Position 1 DOTA2 agents as a result of AWKWARD rearranging the plan for the agent in the Position 5 role. Trend lines represent average over $N=5$ trials.}
    \label{fig:goldOperA}
\end{figure}
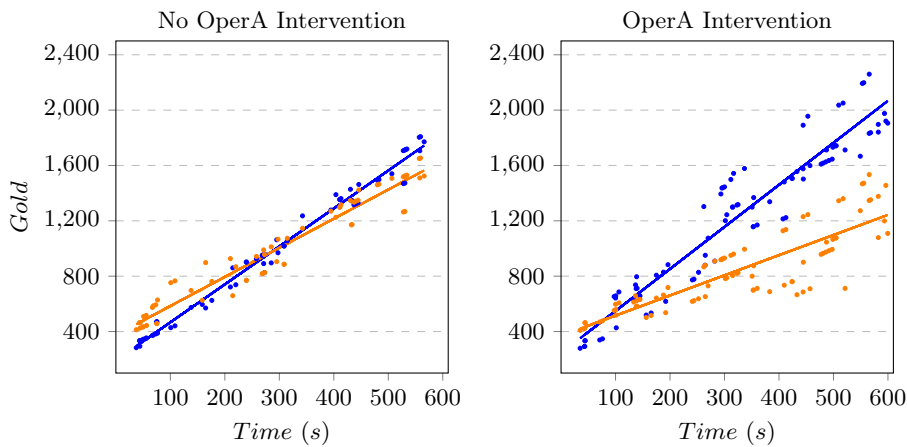

To demonstrate how these changes in the plan impacts the performance of agents, we used gold acquisition as a quantitative metric. This is the standard metric used for players' performance evaluation in DOTA2 tournaments. The agent's value of gold is measured over the course of the first 10 minutes of the game. The right subplot in Figure~\ref{fig:goldOperA} shows the divergence in gold acquisition over time between the AWKWARD bot of position 5 and the default DOTA2 bot of position 1 over five trials. This divergence can be explained by the AWKWARD bot's social behaviour change; OperA banning farm will result in the role sacrificing its own gain and promoting its ally's acquisition of gold instead. The AWKWARD bot's social adjustment is to deny itself from farming when the priority ally (Position 1) is around. 

In this scenario, we mark the moments in time where plan changes are expected to occur due to social interactions, but OperA is not inducing the change in order to see the difference in each bot's gain. While both roles still acquire gold, Position 1 has acquired less than expected, and both agents approximately match each other's gain. 

In contrast, the left subplot in Figure~\ref{fig:goldOperA} illustrates a different trend when the AWKWARD bot does \textit{not} change its plan within the social context. These plots show the two roles on par with one another in terms of gold acquisition over time. When the AWKWARD bot does not alter its plan and continues to attempt to farm, even while the ally of higher priority (Position 1) is also farming in the same lane. The difference in gold acquisition between scenarios can be seen over numerous trials (the dotted lines shows the average linear trend over 5 trials). The data varies in time and gold value due to added randomness across game instances, but the overall trends remain similar.

\section{Related Work}
The AWKWARD architecture combines the normative framework OperA with the Behaviour-Based AI architecture BOD. In this section, we review and compare our architecture to relevant approaches found in the literature. As a complete survey of the literature is beyond the scope of this article, we focus on the most relevant approaches that we considered during the conceptualisation of the AWKWARD architecture.

\subsection{Behaviour-Based AI}
Various approaches of Behaviour-Based AI have been proposed for achieving real-time performance in embodied---physical or virtual---agents including the Subsumption Architecture \cite{brooks86}, Pengi \cite{AgreChapman1987}, and ANA \cite{Maes1990}. These bottom-up reactive planning approaches use condition-action pairs without---or with minimal---internal state; i.e. a simple functional mapping between perceived environmental stimuli and their appropriate responses. Such reactive approaches have proven highly effective for a variety of problems \cite{Mataric1997}. However, in comparison to BOD, these approaches expect little regularity in the arbitration of behaviour; i.e. all possible behaviours must be considered at all times. Moreover, they cannot store information dynamically, thus putting the onus on the developer to predict possible stimuli and develop the appropriate behaviours. BOD overcomes these issues by further decomposing behaviours into ``things that need to be checked regularly, things that only need to be checked in a particular context, and things that one can get by not checking at all'' \cite{Bryson2001}.

While BOD also originated as a robotics cognitive architecture with its POSH implementation \cite{Bryson2001}---and most recently Instinct \cite{Wortham2016Instinct}---it has made its way to virtual environments such as games, e.g. pyPOSH in Unreal Tournament\cite{Brom2006}, POSH-Sharp in StarCraft \cite{GaudlFDG13}, and UNPOSH in Unity \cite{Theodorou2019COG}. However, in these implementations of BOD, there was no mechanism to verify that the agents adhere to their social roles. Instead, unlike our AWKWARD implementation, the plan developer had to ensure that any social norms were accounted for during the plan's development, limiting the possible interactions between agents and team-level performance of the agents. 

\subsection{Normative Agents and Self-Organisation}

The AWKWARD architecture is related to previous work done in the development of \textit{normative agents}. Early work in the literature for normative agents proposes architectures that extend Belief, Desire, Intention (BDI) models with norms \cite{castelfranchi1999deliberative,Kollingbaum2004,lee2014n}. These agents deliberate around the generation and selection of both goals and plans. For instance, \cite{Broersen2002GoalGI} describe their (Beliefs, Obligations, Intentions, and Desires) BOID agent architecture that appends obligations as a mental attitude in addition to its beliefs, desires and intentions. However, as \cite{DignumDignum2020} argue, BDI agents focus towards their own goals instead of social interactions---both with other artificial and human agents---making such approaches unsuitable for multi-agent systems where cooperation between agents is necessary.

Another related approach is N-Jason, a norm-aware BDI agent interpreter equipped with the programming language for agent norm compliance at runtime \cite{lee2014n}. It extends Jason/AgentSpeak(L) with addition of normative concepts such as obligations, permissions, prohibitions, deadlines, priorities, and duration. Similar to N-Jason, N-2APL is also a BDI agent architecture that supports norm-aware deliberation \cite{alechina2012programming}. N-2APL allows agents to adopt normative behaviour in the form of deontic obligations and prohibitions with specified deadlines.
OperA being a framework for formal specifications of social interactions instead of a complete architecture, enables us to define explicitly the social interactions in the form of scene specifications while also keeping the reactive planner component independent; i.e. operational without the OperA module. 

Relevant work also includes existing methodologies for the development of normative agents in multi-agent organisations such as Moise \cite{Hannoun2000MOISEAO} and its extension Moise+ which adds an inheritance on the roles and structural verification features \cite{hubner2007developing,MOISE2002}. Moise+ is based on notions of \textit{roles}, \textit{groups}, and \textit{missions}, enabling explicit specification of MAS organisations that agents can reason about and organisational platforms can enforce \cite{hubner2007developing}. Moise+ offers an implicit description of an interaction protocol through deontic links that specify agent permissions and obligations within their assigned missions. An agent belongs to groups where they are offered a set of permitted roles and missions. Thus, upon changing roles and missions, groups are also subject to change, allowing for task-oriented coalitions to be defined. While interactions in Moise+ are task-driven, OperA leads by social expectation in the form of explicit contracts; a main motivation for its adoption in the work presented here. 

All of these approach capture and represent social norms in order to enable agents to self organise. However, they rely on integrating the social norm enforcement directly into the decision making system. In our architecture, we allow for both reactivity and social deliberation as the situation demands. This ensures that our agents can act efficiently in their environment on their own, i.e. act as complete complex agents, while also organising themselves based on their social roles---and corresponding responsibilities---when they are a part of a larger organisation. Moreover, our decision to use a distributed version of OperA ensures that our AWKWARD agents can interact with other AWKWARD agents, non-AWKWARD agents, and even humans.

Finally, AWKWARD considers that all agents have the capacity to act selfishly and altruistically to varying degrees determined by their roles and social interactions. In contrast to above approaches, we use a reactive planning architecture for the individual agent motivated by its ability to handle uncertainty and dynamic environments. We assign obligations via external expectations that may be subject to change. This approach of global coordination also differs from other work proposed for developing normative reactive planning agents, such as the NoA architecture \cite{Kollingbaum2004}. While NoA adopts norms and deliberates over their activation, our proposed framework concerns itself with dynamically imposing, monitoring and enforcing norms through global coordination, and then distributed enforcement, rather than individual deliberation.

\subsection{Hybrid Approaches with Reactive Planning}
AWKWARD combines formal reasoning, in the form of the OperA module, with reactive planning. A related approach is the logic-based subsumption architecture, where the different control layers that make up a subsumption system have been axiomatised using first-order logic \cite{Amir2001,Amir1998}.
The benefit of this approach is the introduction of non-monotonic reasoning into reactive planning; i.e. developers can understand and easily add control layers to the subsumption planner at run time \cite{Amir2001}. Similarly, the Layered Argumentation System combines---at hardware level---fuzzy reasoning and non-monotonic reasoning for run-time generation of reactive plans \cite{Song2011,Song2006}. These combinations of reactive and formal reasoning approaches bridge together communities---one of our goals---but their focus has been to improve the overall performance of the agent by combining the two paradigms instead of enabling cooperation between multiple agents.

With the latter goal in mind, ABC$^{2}$ architecture combines classical planning with reactive approaches \cite{Matellan2001}. ABC$^{2}$, similar to AWKWARD, emphasises cooperation between agents.
In ABC$^{2}$ each agent has to define and broadcast their own `skills'; i.e. it requires active communication and coordination instead of promoting self-coordination as AWKWARD's OperA implementation does. AWKWARD keeps the reactive and formal parts of the system completely separate from each other; i.e. our reactive planner---sans social consideration---can operate without the OperA module. 

AWKWARD uses both formal reasoning and reactive planning as complementary approaches to each other. However, unlike past approaches, this combination is done on an architectural level. By keeping the reactive planner separate from the normative system, our reactive planner can operate without the normative reasoning module---even if the agent is not behaving within the limits of its socio-organisational role. 
 
\section{Conclusions and Future Work}

In this paper, we presented the AWKWARD architecture for hybrid systems: agents that combine normative reasoning, in the form of OperA, and behaviour-based AI methods, in the form of BOD, at an architectural level. Combining the advantages of BOD and OperA, AWKWARD achieves real-time adjustment of agent plans for evolving social roles as it verifies---and adjust plans to ensure---adherence to any socio-organisational role prescribed to the agent. We provided a toy example, implemented in the game DOTA2, where we demonstrated how AWKWARD enables continual manipulation of agent's behaviour over changing environmental and social circumstances.

With the planner and OperA module implemented in DOTA2, we have demonstrated how OperA can influence the behaviour of reactive planners under defined social circumstances. However, manually designing and building reactive plans can be inefficient and time consuming for system developers, especially as plans scale. As a next step, we intend to investigate methods of optimising plan structure automatically. To do this, Reinforcement Learning techniques can be employed to guide the discovery of an optimal ordering and OpenAI Gym acts as a favourable toolkit for this task. Additionally, we are interested in extending the scope to cover variable autonomy, where varying levels of decision-making control can be passed between a human player---who either directly controls a hero in the game or oversees a team of bots---and artificial agents. 

\subsubsection{Acknowledgements} Theodorou was supported by the Wallenberg AI, Autonomous Systems and Software Program (WASP) funded by the Knut and Alice Wallenberg Foundation. We would like to thank Dignum V. for her input on OperA. All code is available at: https://github.com/lulock/dota

\bibliographystyle{splncs04}
\bibliography{bibliography}
\end{document}